\input{aipcheck}

\documentclass[
    ,final            
  ]
  {aipproc}

\layoutstyle{8x11double}

\begin{document}

\title{MICE:  THE INTERNATIONAL MUON IONIZATION COOLING EXPERIMENT:  PHASE SPACE COOLING 
MEASUREMENT\footnote{Work supported by National Science Foundation Award 757938}}

\classification{29.20.-c, 29.27.Fh, 29.20.db, 14.60.Pq}
\keywords      {muon ionization cooling, neutrino factory, muon collider}

\author{T. L. Hart\footnote{tlh@fnal.gov}}{
  address={University of Mississippi-Oxford, University, MS 38677, U.S.A. \ for the MICE Collaboration}
}



\begin{abstract}
MICE is an experiment with a section of an ionization cooling channel and a muon beam.  
The muons will be produced by the decay of pions from a target dipping into the ISIS proton beam at Rutherford Appleton Laboratory (RAL).  
The channel includes liquid-hydrogen absorbers providing transverse and longitudinal momentum loss and high-gradient radiofrequency (RF) cavities for longitudinal reacceleration,
all packed into a solenoidal magnetic channel.  MICE will reduce the beam transverse emittance by about 10\% for muon momenta 
between 140 and 240 MeV/c.  Time-of-flight (TOF) counters, a threshold Cherenkov counter, and a calorimeter will identify background 
electrons and pions.  Spectrometers before and after the cooling section will measure the beam transmission and input and output 
emittances with an absolute precision of 0.1\%.   
\end{abstract}

\maketitle


\section{Introduction}
A Neutrino Factory based on a muon storage ring is the ideal tool for neutrino oscillation studies and possibly for the discovery of leptonic CP violation.  A Neutrino Factory could also be the first step toward a $\mu^+ \mu^-$ collider.  Ionization cooling, while not yet demonstrated, has been shown by simulations and design studies to be an important factor for the performance and cost of a Neutrino Factory.  An international R\&D program for the development of a Neutrino Factory [1] and $\mu^+ \mu^-$ collider [2] has been established.  An important step toward these facilities is a first experimental demonstration of muon ionization cooling.  The main goals of the international Muon Ionization Cooling Experiment [3] are to:

\begin{list}{\labelitemi}{\leftmargin=1em}
\item
{Design, engineer, and build a section of cooling channel with performance suitable for a Neutrino Factory,}

\item
Run the cooling channel in a muon beam, and measure its performance in various modes and beam conditions to test the limits 
and practicality of muon cooling.
\end{list}

\section{Experimental Layout}

The main components of MICE are shown in Fig. 1.  Cooling is provided by one cell from the 2.75 m cooling channel of ``Study-II'' [4].  
Some components of the Study-II cooling channel have been modified to reduce costs and to comply with RAL safety requirements.  
The incoming muon beam first encounters a TOF counter, a Cherenkov detector, and then a second TOF counter~[5].  The 
TOF counters make precise time measurements which contribute to particle identification (PID).  Further PID is done with the Cherenkov 
detector to reject pion backgrounds in the muon beam.  
Next,
a lead diffuser generates a tunable 
input emittance.  Then a spectrometer with tracking detectors within a uniform solenoidal magnetic field measures the locations and 
momenta of each particle.  After this initial momentum measurement is the cooling section consisting of liquid hydrogen absorbers, RF cavities [6], 
and superconducting coils.  An additional absorber finishes the cooling section to protect the downstream tracker from dark currents emitted by 
the RF cavities.  The track positions and momenta after the cooling section are measured by a second spectrometer identical to the 
first,  
plus
a calorimeter and third TOF counter to provide further time and PID measurements and reject background electrons from muon decay.

\begin{figure}
  \includegraphics[width=.75\textwidth]{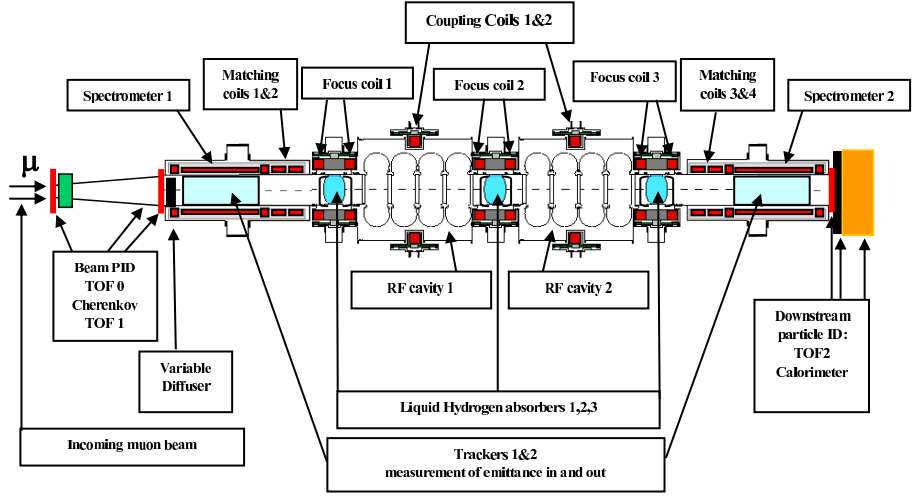}
  \caption{MICE layout}
\end{figure}

\section{Beam Emittance Measurement}

The emittance of a beam is an approximation to the phase space volume occupied by the particles comprising the beam.  
Using a coordinate system in which $z$ is along the beam direction, the 6-dimensional normalized emittance is the 6$^{\rm{th}}$ root of 
the determinant of the covariance matrix of ($t,$ $x,$ $y,$ $E,$ $p_x,$ $p_y$) divided by the quantity muon mass times the speed of light.  The 
4-dimensional normalized emittance, or normalized transverse emittance, is the 4$^{\rm{th}}$ root of the determinant of the covariance 
matrix of ($x,$ $y,$ $p_x,$ $p_y$) divided by the quantity muon mass times the speed of light [7].  Defining the covariance 
matrices $(M_{ij})_{6D} = \sigma(u_i, u_j)_{6D}$ where $u_{\,6D}$ = ($t,$ $x,$ $y,$ $E,$ $p_x,$ $p_y$) and 
$(M_{ij})_{4D} = \sigma(u_i, u_j)_{4D}$ where $u_{\,4D}$ = ($x,$ $y,$ $p_x,$ $p_y$),
 
 \begin{equation}
 \epsilon_N^{6D} =  \frac{\sqrt[6]{{\rule{0pt}{9pt}\rm{det}}(M_{ij})_{6D}}}{m_{\mu}c}, \quad
 \epsilon_N^{4D} =  \frac{\sqrt[4]{{\rule{0pt}{9pt}\rm{det}}(M_{ij})_{4D}}}{m_{\mu}c} 
 \end{equation}
 
 The rate of change of the 4-dimensional normalized emittance [8] is given by

\begin{equation}
\frac{\rule[-4pt]{0pt}{3pt}d\epsilon_N^{4D}} {dz} 
= - \frac{1}{\beta^2} \left| \frac{dE_{\mu}}{dz}\right| \frac{\rule[-4pt]{0pt}{3pt}\epsilon_N^{4D}}{E_{\mu}}
+ \frac{\beta_{\perp}(0.014 \,{\rm{GeV}})^2}{2 \, \beta^3 E_{\mu} \, m_{\mu} \, X_0} \  \  {\rm{where}}
\end{equation}


\begin{list}{\labelitemi}{\leftmargin=1em}

\item
$X_0$ is the radiation length of the material through which the muons are passing,

\item
$\beta$ = v/c, the muon velocity divided by that of light,

\item
$\beta_{\perp}$  is the beta function of the magnetic system the muons are passing through,

\item

$| \, dE_{\mu} / dz \, |$ is the mean rate of energy loss of the muons through the material.
\end{list}

Setting $d\epsilon_N^{\,4D}\!/{dz} = 0$ yields the equilibrium emittance

\begin{equation}
\epsilon_{N, \, eq}^{\,4D} = \frac{\beta_{\perp}(0.014 \,{\rm{GeV}})^2}{2 \, \beta E_{\mu} \, m_{\mu} \, X_0 \, | dE_{\mu} / dx |}
\end{equation}

To make the equilibrium emittance small, the beta function must be small and the rate of energy loss and radiation length of the absorber material must be large.  These motivate the choices of a strong focusing magnetic system and liquid hydrogen for the absorber material.  For MICE, which uses liquid hydrogen absorbers, the equilibrium emittance is about 2.5$\pi$ mm radians.

\section{Measurement Technique}

Precision measurements of muon beam transmission and emittance require single particle tracking and detection using standard particle physics techniques instead of those typically used in beam instrumentation.  Momentum measurements are made with magnetic spectrometers.  Each spectrometer measures x and y coordinates of an incident particle at given z positions.  Momentum and angles are reconstructed by fitting a helix.  The root mean square resolution of the position measurements must be less than about 10\% of the root mean square beam size for the experimental resolution to remain small compared to the emittance measurement resolution.  A precision measurement of the emittance also requires tracking that can determine if a particle left the MICE channel or completely traversed it, in order to separate the effects of beam loss and cooling.  The MICE PID system needs to keep the electron and pion contamination of the muon beam below 0.1\% to achieve the desired emittance measurement precision.  Measurements of accelerator and absorber parameters (RF phase and voltages, absorber thickness, etc.) are also needed for consistency checks and to ensure measurement reproducibility. 

\section{Detectors}

The main criteria for the MICE detector are precision, robustness of the tracking detectors in the potentially severe backgrounds of the RF cavities, and redundancy in PID to keep contamination below 0.1\%.  The two upstream TOF detectors consist of 6 cm wide hodoscope layers segmented in x and y directions, and the downstream TOF detector consists of 4 cm wide hodoscope layers also segmented in x and y.  The timing resolution of the TOF detectors is about 60 ps as measured in a test beam.  The upstream and downstream spectrometers each contain five sets of scintillating fiber planes deployed in three stereo views.  Groups of seven fibers are read out using cryogenic VLPC photodetectors.  The threshold Cherenkov system upstream of the cooling channel is two detectors containing aerogel radiators.  The aerogel indices of refraction are 1.07 and 1.12.  Each is read out with four photomultiplier tubes.  The indices of refraction are such that over track momenta between 140 and 240 MeV/c, pions and muons going through the two detectors can be distinguished by the intensity of the Cherenkov light in each detector.  The downstream PID system has  a third TOF counter and an electron/muon ranger.  The 
calorimeter consists of two detector systems:  a layer of lead with embedded scintillating fibers used to detect electrons, and ten-layers of scintillator planes to detect (penetrating) muons.  Careful diagnostic design and attention to system integration and calibration will enable emittance measurements with 0.1\% precision.

\section{Status and Immediate Plans}

    The MICE Collaboration, roughly 130 physicists and engineers from the world's accelerator and particle physics communities, will soon be bringing together the MICE detector and cooling channel components at RAL.  Two TOFs, a Cherenkov counter, and two tracker systems have been assembled.  The TOFs and Cherenkov counter have been placed in a preliminary beam, and the tracker systems have been calibrated with data from cosmic rays.  The beam line up to the diffuser is completed.  Initial emittance measurements are expected after the installation of the spectrometer solenoid later this year.  The sequential addition of the second tracker, absorbers, and RF cavities will culminate in extensive emittance and muon cooling measurements in varying run conditions.  

    An example of a set of emittance measurements that will be made at MICE is shown in Figure 2, which shows a GEANT-based simulation of the fractional change of transverse emittance before and after the cooling channel for various initial emittances.  A nominal initial emittance of 6$\pi$ mm radians allows a 10\% emittance reduction.  The simulation verifies that the reconstructed equilibrium emittance, at which the fractional emittance change is zero, is consistent with the calculated value of  2.5$\pi$ mm radians.  

       MICE will operate with a variety of settings allowing cooling performance to be mapped out for a range of cooling-channel parameters and beam momenta.  The cooling performance will be compared with predictions of detailed simulations.  A demonstration that muon ionization technology is feasible and 
that its cost and performance are well understood will pave the way for the Neutrino Factory Conceptual Design Report and will provide direction to muon colliders in the longer term.

\begin{figure}[h!]
  \includegraphics[width= 0.95\linewidth]{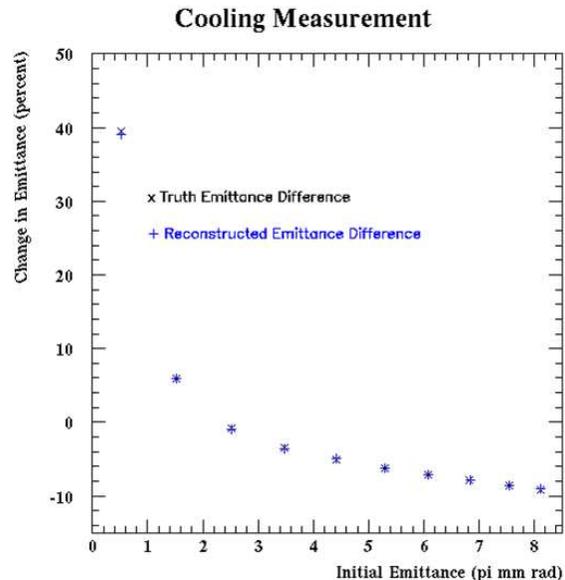}
  \caption{GEANT-based simulation of fractional emittance change vs.\ initial emittance (nominally 6$\pi$ mm radiams) of a muon beam going through the MICE cooling channel.}
\end{figure}



%





\end{document}